\documentclass[pre,twocolumn,showpacs]{revtex4}
\usepackage{graphicx}
\usepackage{amsmath}
\begin{document}
\title{Phase transition in a triplet process}
\author{Kwangho Park$^{1,3}$, Haye Hinrichsen$^{2}$, and In-mook Kim$^{1}$}
\affiliation{$^{1}$Department of Physics, Korea University, Seoul,
        136-701, Korea}
\affiliation{$^{2}$ Theoretische Physik, Fachbereich 8, Universit{\"a}t
        GH Wuppertal, 42097 Wuppertal, Germany}
\affiliation{$^{3}$ Fakult\"at 4, Theoretische Physik,
        Gerhard-Mercator-Universit{\"a}t Duisburg,
        47048 Duisburg, Germany}
\date{\today}
\begin{abstract}
We argue that the reaction-diffusion process $3A \to 4A, \; 3A \to 2A$
exhibits a different type of continuous phase transition from an
active into an absorbing phase. Because of the upper critical
dimension 
$d_c\geq 4/3$ we expect the phase transition in 1+1 dimensions to be  characterized by nontrivial fluctuation effects.
\end{abstract}
\pacs{05.70.Ln, 64.60.Ak, 64.60.Ht}
\maketitle
\def\xvec{\text{\bf x}}
\parskip 1mm 

The classification of continuous phase transitions far from thermal equilibrium is one of the most challenging tasks of modern statistical physics~\cite{MarroDickman}. Within this field many studies are concerned with phase transitions from a fluctuating active phase into one or several nonfluctuating absorbing states, which are believed to be associated with a finite number of universality classes~\cite{Hinrichsen}. For such a phase transition to occur it is necessary that (a) at least one absorbing state is dynamically accessible, (b) there are two competing processes for particle creation and removal, and (c) there is a mechanism which prevents the particle density from diverging. 

In many cases, the critical behavior close to the transition is
characterized by simple power law scaling. In sufficiently high
dimensions the critical exponents are given by their mean-field
values, whereas below a certain upper critical dimension $d_c$
fluctuation corrections have to be taken into account, leading to
nontrivial exponents and scaling functions. For this reason the study
of fluctuation effects in low dimensional, especially 1+1-dimensional systems is particularly interesting.

The most important universality class of absorbing phase transitions is directed percolation (DP), which occurs in all processes following the reaction-diffusion scheme $A \leftrightarrow 2A, A \to \emptyset$. The critical exponents, especially in one spatial dimension, are known to a very high precision~\cite{Jensen}. The critical behavior of DP can be described in terms of a renormalizible field theory which was originally introduced in the context of high energy physics~\cite{Reggeon}. DP is relevant for experimental applications such as catalytic reactions~\cite{ZGB86}, flowing sand~\cite{HJRD99}, and spatiotemporal intermittency of magnetic fluids~\cite{RRR02}. 

The other established class is the parity-conserving (PC) universality class. This type of critical behavior is observed in a large variety of models which can be divided into two groups. The first group includes all parity-conserving particle processes~\cite{ABModel,TakayasuTretyakov92,ALR93,CardyTauber96} such as branching-annihilating random walks with two offspring $A \to 3A, 2A \to  \emptyset$. The second group of models comprises spreading processes with two symmetric absorbing states, including kinetic Ising models~\cite{Menyhard94}, interacting monomer-dimer models~\cite{KimPark94}, as well as generalized versions of the  Domany-Kinzel model and the contact process~\cite{Hinrichsen97}. In higher dimensions the second group of models describes branching-annihilating {\it interfaces} and can be associated with the voter universality class~\cite{Voter}. Only in 1+1 dimensions the two classes of models exhibit the same type of critical behavior. 

Recently the pair contact process with diffusion (PCPD), also called annihilation-fission process, attracted considerable attention. The PCPD is a {\it binary} spreading process following the reaction-diffusion scheme
\begin{equation}
\label{RDScheme}
nA \to (n+1)A \,, \qquad nA \to mA ,
\end{equation}
with $n=2,m\le 1$. It exhibits a continuous phase transition and thus
could serve as a candidate for another independent universality
class. The PCPD was already suggested in 1982 by
Grassberger~\cite{Grassberger82}, but it took almost 20 years until
Howard and T\"auber presented a first systematic study of a bosonic
variant of the process~\cite{HowardTauber97}. Using field-theoretic
methods they were able to prove the existence of a phase transition,
although the corresponding field theory turned out to be
unrenormalizable. More recently, several authors studied various
fermionic variants of the
PCPD~\cite{Carlon,Hinrichsen1,Odor,Coag,CyclicPaper,PreviousPaper,NohPark,OMS02}.
Meanwhile there is a general consensus that the critical behavior of
the PCPD is different from all other previously known universality
classes. However, it turned out to be extremely difficult to estimate
the critical exponents in a reliable way, mainly because of unusually
strong deviations from ordinary power-law scaling. 

In the present study we investigate the question whether further different
types of critical behavior will emerge for $n>2$.
In particular we will focus on the case $n=3$, called triplet
process. As will be shown below, we argue that this process exhibits
yet another different type of critical behavior. 

\begin{figure}
\includegraphics[width=75mm]{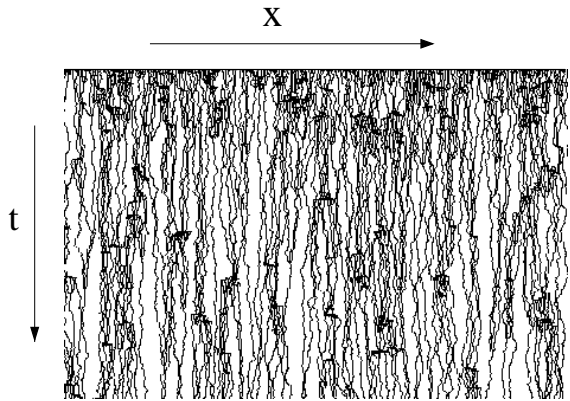}
\caption{
\label{FIGTPDEMO}
One-dimensional triplet process at criticality starting with
a fully occupied initial state.
}
\end{figure}

\paragraph{Mean field approximation.}
In order to determine critical dimension $d_c$ and the mean field
critical exponents  of the reaction-diffusion process
(\ref{RDScheme}) for general $m<n$,
let us consider a simple mean field theory.
We expect this process to be described by the Langevin equation
\begin{equation}
\label{Langevin}
\partial_t  \rho(\xvec,t) = a\rho^n(\xvec,t) - 
\rho^{{n+1}}(\xvec,t)+D\nabla^2\rho(\xvec,t) + \zeta(\xvec,t)\,,
\end{equation}
which for $n=1$ reduces to the well-known Langevin equation for
DP~\cite{DPConjecture}. The first term accounts for both particle
creation and removal so that the parameter $a$ plays the role of the
reduced spreading probability $p-p_c$. The second term is the most
relevant contribution preventing the particle density from going to
infinity, while the third term describes nearest-neighbor
diffusion.

The noise $\zeta(\xvec,t)$ takes the stochastic nature of
particle creation and removal into account. Its amplitude 
has to depend on the local density $\rho(\xvec,t)$ since in the absorbing state $\rho=0$ there are no density fluctuations. Thus it is near at hand to expect  noise correlations of the form
\begin{equation}
\label{Noise}
\langle \zeta(\xvec,t) \zeta(\xvec',t') \rangle = 
\Gamma \rho^\mu (\xvec,t) \delta^d (\xvec-\xvec') \delta(t-t')\,
\end{equation}
with an unknown exponent $\mu$. For DP ($n=1$), where the squared noise amplitude is proportional to the density of particles, this exponent is given by $\mu=1$. For $n>1$, however, the situation is more involved. Without the branching process, i.e., deep in the inactive phase, the squared noise amplitude is proportional to $\partial_t \rho(\xvec,t)$, hence $\mu=n$. At the transition, however, the branching process may lead to positive correlations among the particles, increasing the intensity of the noise and thereby reducing the value of $\mu$. At criticality we therefore expect $\mu$ to be in the range 
\begin{equation}
1 \leq \mu \leq n.
\end{equation}
Solving the stationary mean field equation $0=a\rho^n - \rho^{{n+1}}$ we obtain the stationary density $\rho=a$, hence the critical point is $a_c=0$ and the density exponent is $\beta^{\rm MF}=1$. At the mean field critical point the full Langevin equation should be invariant under the scaling transformation
\begin{equation}
\label{Rescaling}
\xvec \to \Lambda \xvec
\,, \quad
t \to \Lambda^z t
\,, \quad
\rho \to \Lambda^{-\chi} \rho \,,
\end{equation}
where $\Lambda$ is a dilatation factor while
$z=\nu_\parallel/\nu_\perp$ and $\chi=\beta/\nu_\perp$ are quotients
of the three standard critical exponents.  Comparing all terms except for the noise, scaling invariance implies that $z=2$ and $\chi=2/n$, i.e.,
\begin{equation}
\label{MFExponents}
\beta^{\rm MF}=1, \quad 
\nu_\perp^{\rm MF}=n/2, \quad
\nu_\parallel^{\rm MF}=n \,.
\end{equation}
Moreover, we can check the relevance of the noise term, which is responsible for fluctuation effects. By simple power counting we find that the noise is relevant below the upper critical dimension 
\begin{equation} 
d_c=2+\frac{4-2\mu}{n} \,,
\end{equation}
while it is irrelevant above $d_c$ where the mean field exponents~(\ref{MFExponents}) are expected to become exact. For DP ($n=\mu=1$) we obtain the well-known result $d_c=4$, while for the PCPD the upper critical dimension has to be in the range $2\leq d_c\leq 3$. This result is in agreement with recent numerical findings by \'Odor {\it et al.}~\cite{OMS02} suggesting that $d_c=2$.

\begin{figure}
\includegraphics[width=75mm]{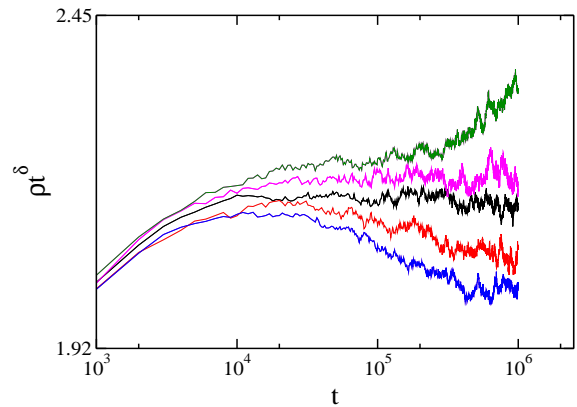}
\caption{
\label{DELTA}
The density of particles, $\rho(t)$, times $t^\delta ~(\delta=0.32)$ as a function of
time for $p=0.6855, 0.6853, 0.6851, 0.6849$, and
$0.6847$ from top to
bottom, averaged over $1500$ runs on a system with $4096$ sites.
The best straight line is obtained for $\delta=0.32$ and $p=0.6851$.
}
\end{figure}

As the main observation, which triggered the present work, we note
that the upper critical dimension for third-order processes 
($n=3$, $m<n$, $1\leq\mu\leq3$) is larger than $4/3$. Consequently, in 1+1 dimensions such a triplet process (TP) should still be characterized by non-trivial fluctuation effects. Moreover, the density in the inactive phase is known to decay as $\rho(t) \sim (\ln t/t)^{1/2}$.  This type of decay in the absorbing phase differs from all other previously  known universality classes of phase transitions into absorbing states, suggesting that also the transition itself should belong to yet another universality class.

\paragraph{Numerical simulations.}
In order to study the triplet process numerically, we introduce a
fermionic variant of the reaction-diffusion process~(\ref{RDScheme}) 
with $n=3$ and $m=2$. It evolves by random-sequential updates and is defined by the following dynamic rules:
\begin{equation}
\begin{array}{ccr}
\emptyset A \rightarrow A \emptyset & \text{ with rate } & (1-p)/2\\
A \emptyset \rightarrow \emptyset A &  & (1-p)/2\\
AAA \to AA\emptyset                     &  & (1-p)/2 \\
AAA \to \emptyset AA                    &  & (1-p)/2 \\
AAA\emptyset  \to AAAA                  &  & p/2 \\
\emptyset AAA  \to AAAA                 &  & p/2 
\end{array}
\end{equation}
A typicial space-time plot of the process at criticality is shown in Fig.~\ref{FIGTPDEMO}. As can be seen the process generates
spatiotemporal structures, possibly indicating the presence of fluctuation effects. Note that we tuned the rates for diffusion and particle removal in the same way as in Ref.~\cite{PreviousPaper}.

Performing standard Monte Carlo simulations (see,
e.g.,~\cite{Hinrichsen}) we find clear evidence for a continuous phase
transition between an active phase, where the denstity of particles is
asymptotically constant, and an inactive phase, where the particle density
decays algebraically with logarithmic corrections. Assuming that the critical
behavior at the transition obeys simple power law scaling we find the
critical threshold $p_c=0.6851(4)$(see Fig.~\ref{DELTA}). 
As in the PCPD, there are strong corrections so that the scaling
regime, if existent at all, is not reached before $10^4$ time steps. 
Averaging over many independent runs in the time interval $10^4 < t < 10^6$ we estimate the critical exponents by
\begin{equation}
\begin{array}{ccr}
\nu_\parallel = 2.5(2), &  z=1.75(10), & \delta= {\beta}/{\nu_\parallel} = 0.32(1). 
\end{array} 
\end{equation}
Similar exponents were obtained in other variants of the triplet
process with $m<n=3$ (not reported here). In all cases the estimates differ from the mean
field (MF) exponents $\nu^{\rm MF}_\parallel = 3$, $z^{\rm MF}=2, $ and $\delta^{\rm MF}=1/3$, leading us to the
conclusion that critical behavior of the 1+1-dimensional TP is indeed
characterized by non-trivial fluctuation effects. As expected, these
deviations are quite small (less than $20\%$) since the simulations
are carried out close to the upper critical dimension. 

We note that  our results are not accurate enough to doubtless confirm
the validity of power-law scaling over a large range. 
As our simulations seem to reach the scaling regime only after $10^4$
time steps, the accuary of our estimate for $p_c$ is limited as well. 
In addition, the asymptotic power-law behavior may be shadowed by
logarithmic corrections which are also present in the inactive
phase. Finally, as in the case of the PCPD, the assupmtion of simple
power-law scaling and the concept of universality may be questioned as
a whole. 
Nevertheless we believe that the MF arguments and the numerical
evidence are
 strong enough to conclude that this model exhibits a different type of critical behavior, where fluctuation effects are likely to play an important role.

Regarding the limited accuracy of numerical simulations a major drawback could be the definition of the model as a fermionic reaction-diffusion process with four-site interactions. A bosonic variant with two-site interactions is currently under investigation. Moreover, it is important to determine the exponent $\mu$ in the noise correlator. Preliminary simulations suggest a value close to $\mu\approx 2$. Finally, the influence of the diffusion rate has to be studied systematically.

We would like to thank G. \'Odor for numerous helpful discussions.
This work was supported in part by the Korea Research Foundation 
Grant No. KRF-2001-015-DP0120 and also in part
by the Ministry of Education through the BK21 project.


\end{document}